\documentclass[sn-nature]{sn-jnl}

\usepackage{graphicx}
\usepackage{caption}
\usepackage{multirow}
\usepackage{mathrsfs}
\usepackage[title]{appendix}
\usepackage{xcolor}
\usepackage{textcomp}
\usepackage{manyfoot}
\usepackage{booktabs}
\usepackage[switch,columnwise]{lineno}
\usepackage{lipsum}
\usepackage{lettrine}
\usepackage{threeparttable}
\usepackage{xcolor}
\usepackage{siunitx}
\DeclareSIUnit{\bit}{bit}
\usepackage{amsmath,amssymb,amsfonts,amsthm}
\usepackage{physics}
\usepackage{xfrac}
\usepackage{nicefrac}
\usepackage{bm}
\usepackage{mathtools}
\usepackage[utf8]{inputenc}
\usepackage{booktabs}
\usepackage{arydshln}
\usepackage{array}
\usepackage{stackengine}
\usepackage{ragged2e}
\usepackage{microtype}
\usepackage[]{hyperref}
\hypersetup{
	pdfauthor={Thomas Daunizeau},
	pdfcreator={Thomas Daunizeau},
	pdfproducer={Thomas Daunizeau},
	colorlinks,
	citecolor=black,
	filecolor=black,
	linkcolor=black,
	urlcolor=black
}
\usepackage{layouts}
\justifying

\renewcommand{\figurename}{\scriptsize{Fig.}}

\newcommand{\minus}{\scalebox{0.70}[1.0]{$-$}}

\unnumbered
\begin{document}

\title[Article Title]{\centering Dynamic Shaping of Multi-Touch Stimuli by\\Programmable Acoustic Metamaterial}

\author*[]{\fnm{Thomas} \sur{Daunizeau$^{\ast}$}}\email{\href{mailto:thomas.daunizeau@sorbonne-universite.fr}{thomas.daunizeau@sorbonne-universite.fr}}
\author[]{\fnm{Sinan} \sur{Haliyo}}
\author[]{\fnm{David} \sur{Gueorguiev}}
\equalcont{}
\author[]{\& \fnm{Vincent} \sur{Hayward$^{\S}$}}
\equalcont{These authors contributed equally to this work.\quad $^{\S}$Deceased.\phantom{\hspace{29.95mm}}}
\affil[]{\orgdiv{Sorbonne Université, CNRS, ISIR}, \postcode{F-75005} \city{Paris}, \country{France}}

\abstract{\textbf{Acoustic metamaterials are artificial structures, often lattice of resonators, with unusual properties. They can be engineered to stop wave propagation in specific frequency bands. Once manufactured, their dispersive qualities remain invariant in time and space, limiting their practical use. Actively tuned arrangements have received growing interest to address this issue. Here, we introduce a new class of active metamaterial made from dual-state unit cells, either vibration sources when powered or passive resonators when left disconnected. They possess self-tuning capabilities, enabling deep subwavelength band gaps to automatically match the carrier signal of powered cells, typically around \SI{200}{\hertz}. Swift electronic commutations between both states establish the basis for real-time reconfiguration of waveguides and shaping of vibration patterns. A series of experiments highlight how these tailored acceleration fields can spatially encode information relevant to human touch. This novel metamaterial can readily be made using off-the-shelf smartphone vibration motors, paving the way for a widespread adoption of multi-touch tactile displays.}}

\keywords{Acoustic metamaterials, Programmable matter, Haptics}

\maketitle

\section{Introduction}

The control of wave propagation lies at the foundation of numerous applications, for which acoustic metamaterials hold immense potential~\cite{LuEtAl-09, MaSheng-16}. They are engineered structures, often lattice of subwavelength resonators~\cite{LiuEtAl-00, HuangSun-10}. Their collective action endows the bulk material with unusual effective properties such as a negative mass density and/or a negative modulus~\cite{HuangEtAl-09, LeeEtAl-09, LeeEtAl-10, BrunetEtAl-15}. These properties, typically not found in nature, unlock novel wave phenomena~\cite{CummerSchurig-07, Pendry-00, LemoultEtAl-11}, including the ability to stop wave propagation in specific frequency ranges called band gaps. Unlike those derived from Bragg scattering~\cite{Martinez-SalaEtAl-95}, these band gaps stem from the hybridization of local resonances with free-space dispersion. Crucially, they are independent of wavelength, allowing for compact and practical metamaterial designs.

Following their success in optics~\cite{JoannopoulosEtAl-97} and acoustics~\cite{CummerEtAl-16}, metamaterials have sparked interest in haptics~\cite{BilalEtAl-20}. Unlike conventional control and geometric methods~\cite{PanteraHudin-20, ReardonEtAl-23, DhiabHudin-20}, they are robust, versatile, and highly effective at confining elastic waves for the rendering of precise, vibration-based tactile feedback~\cite{DaunizeauEtAl-21}. However, once manufactured, these passive structures possess fixed dispersive properties and spatial arrangements that limit their use, whether in haptics or other fields. Additionally, their operating frequency is bounded by the Kramers-Kronig relations~\cite{Mangulis-64}.

To address these issues, researchers have introduced active unit cells~\cite{Zangeneh-NejadFleury-19}. They support various functions including loss compensation in non-Hermitian metamaterials~\cite{RivetEtAl-18}, reconfiguration of waveguides~\cite{LiEtAl-19a}, and adjustment of band gaps~\cite{ChenEtAl-14}. Piezoelectric transducers are widely used for activating these cells, mainly due to their adjustable stiffness through shunting. They were leveraged to alter resonances by making variable junctions~\cite{BergaminiEtAl-14, ChenEtAl-16a}, establish tunable acoustic filters~\cite{BacigalupoEtAl-20}, trap waves spatially following a specified spectrum~\cite{AlshaqaqEtAl-22}, and reconfigure acoustic lenses in real time~\cite{PopaEtAl-15}. Such dynamic systems are best suited for low-amplitude, high-frequency applications, typically in the \SI{}{\kilo\hertz} to \SI{}{\mega\hertz} ranges. Electromagnetic actuation offers complementary capabilities. It often involves moving an extraneous permanent magnet to remotely toggle the effective modulus between positive and negative~\cite{YuEtAl-18}, or selectively activate resonators for outlining waveguides~\cite{BilalEtAl-17, LeeEtAl-20}, albeit with a \SI{300}{\milli\second} latency per cell. Alternatively, electromagnetic actuation can be distributed within each unit cell by embedding energized coils. They can set bistable or tristable resonators in distinct states, thus controlling band gaps~\cite{WangEtAl-16, YangEtAl-17}, phase changes~\cite{MaEtAl-18}, and polarization~\cite{LiuEtAl-19}. This approach also achieves low-frequency band gaps under \SI{100}{\hertz} via continuously variable stiffness adjustments~\cite{WangEtAl-20}. 

By enabling real-time reconfiguration of waveguides, active metamaterials could be the key to developing a multi-touch, programmable vibration-based display, a major goal of haptic research~\cite{BasdoganEtAl-20}. Human touch is exquisitely sensitive to rapid transients and low-frequency vibrations up to about \SI{1}{\kilo\hertz}~\cite{JohanssonEtAl-82, BolanowskiEtAl-88}, requiring deep subwavelength unit cells the size of a finger pad. However, current active unit cells do not meet haptic requirements. Electromagnetic-based cells are not configured for real-time response, while piezoelectric-based cells operate at excessively high frequencies. Additionally, existing active unit cells can only control a single global vibration source, such as ambient noise, rather than multiple stimuli.

\begin{figure}[!b]
\centering
\includegraphics[width=180mm]{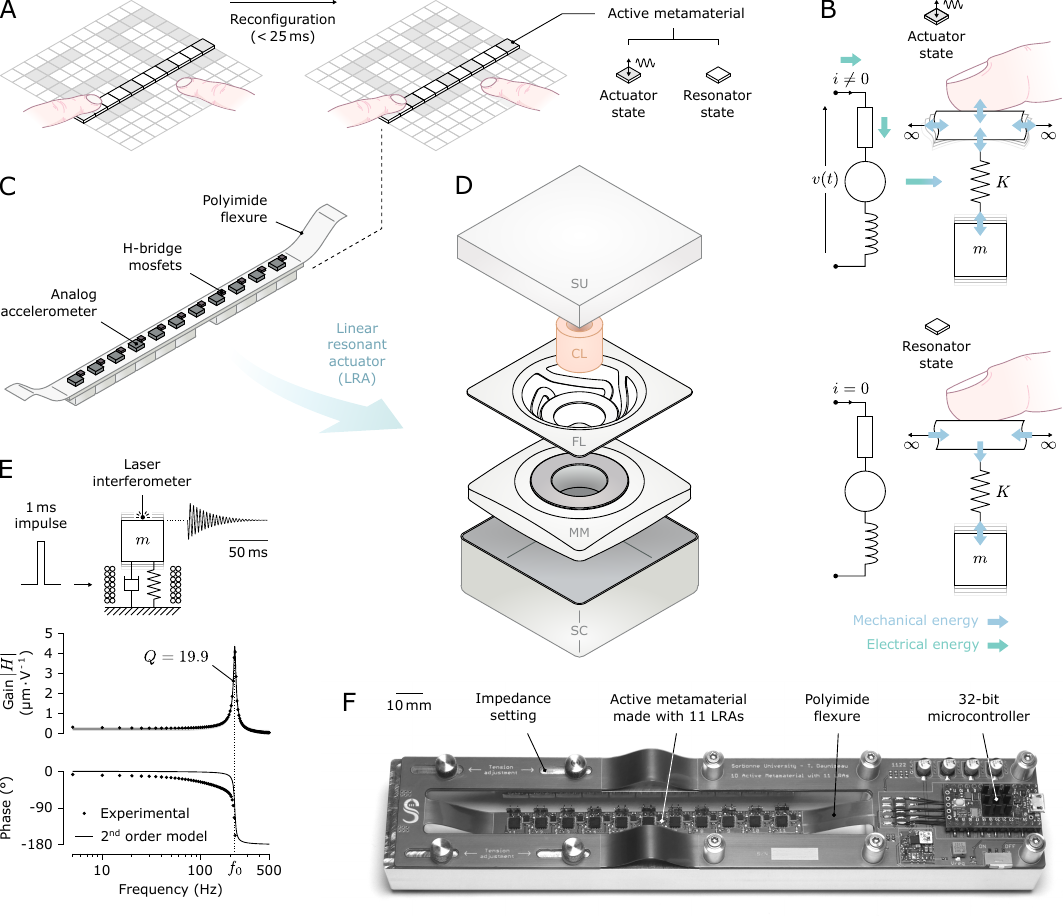}
\caption{\scriptsize{\textbf{Overview of the dual-state active acoustic metamaterial.} \textbf{A.} Schematic of a 2D tactile display made from a square tessellation of dual-state unit cells. Cells in a resonator state (blank) form a metamaterial insulating cells in an actuator state (colored). The resulting vibration patterns can be reconfigured in real time. A reduced linear array of 11 unit cells is outlined in black. \textbf{B.} Equivalent electromechanical model given in both states: the ``actuator state'' and the ``resonator state''. Mechanical and electrical energy flux are overlaid (dissipative phenomena are discarded). \textbf{C.} Schematic of the linear array of 11 unit cells made with LRAs. \textbf{D.} Exploded view of an off-the-shelf LRA with, from top to bottom, a FR-4 substrate (SU), a voice coil (CL), a flexure spring (FL), a moving mass with a neodymium magnet (MM), and a steel casing (SC). \textbf{E.} Impulse response of a single LRA measured using a laser interferometer. Both the gain and phase, averaged over 50 trials, are well approximated by a second order model with minimal damping. \textbf{F.} Top view of the prototype with embedded electronics.}}
\label{fig:overview}
\end{figure}

To overcome these limitations, we introduce a novel type of active acoustic metamaterial. It consists of dual-state unit cells made from off-the-shelf electromagnetic linear resonant actuators (LRAs), commonly found in smartphones and game controllers. Both theoretical and experimental results show that a 1D array of LRAs induces a self-tuned, deep sub-wavelength band gap, within touch-relevant frequencies. We report a method to tailor acceleration fields in real-time, enabling the spatial encoding of binary words and time-varying patterns. We demonstrate its relevance for tactile perception. This work opens up new avenues for responsive multi-touch haptic displays, mechanical computing, and overall democratizes active metamaterials through a low-cost, non-expert approach.

\section{Results}

\subsection{Dual-state active unit cell}

An off-the-shelf LRA with a square footprint of $10\! \times \!10\! \times \!\SI{4}{\milli\meter\cubed}$ (VLV101040A, Vybronics) was chosen to make subwavelength unit cells that tile the plane, as shown in Fig.~\ref{fig:overview}.A. This resonant actuator, illustrated in Fig.~\ref{fig:overview}.D, consists of a mass attached to a spiral flexure spring, which restricted oscillations to the $z$-axis. Requisite active elements are already embedded inside typical LRAs, including a neodymium magnet attached to the moving mass and a coil fixed to the casing. When an alternating current is applied to the coil, the Lorentz force causes the mass to oscillate, generating transverse waves in the base that produce vivid vibrotactile sensations. This was termed the ``actuator state''\!, as schematized in Fig.~\ref{fig:overview}.B. On the other hand, if the coil is left disconnected, i.e. in an open circuit, no current can establish and the unit cell remains a passive resonator. This will further be referred to as the ``resonator state''\!. The possibility of a third state that would allow unimpeded vibration propagation was also investigated. This involved shorting the coil to induce eddy current damping in an attempt to nullify resonances. However, it proved insufficient, achieving only a 10.5\% reduction in quality factor (see Supplementary Fig.~1).

The dynamic response of a unit cell, central to the formation of a band gap, was assessed experimentally (see Methods). As shown in Fig.~\ref{fig:overview}.E, the results are in excellent agreement with a second-order linear time-invariant model ($\forall \, f \in [0,\SI{500}{\hertz}],\, \mathrm{RMSE}=\SI[inter-unit-product = \!\cdot\!]{47}{\nano\meter\per\volt}$\!, see Supplementary Information). The moving mass, $m=\SI{1.57}{\gram}$, combined with a compliant spring, $K=\SI[inter-unit-product = \!\cdot\!]{3.08}{\newton\per\milli\meter}$\!, gave a resonance frequency $f_0=\omega_0/2\pi=\SI{223}{\hertz}$. The DC gain, $G$, was evaluated at $\SI[inter-unit-product = \!\cdot\!]{218}{\nano\meter\per\volt}$\!. The high quality factor, $Q=1/2\,\zeta=19.9$, achieved through frictionless guiding and minimal damping, is ideal for making a resonant acoustic metamaterial. For simplicity, the close-packed square lattice was reduced to a linear arrangement, as illustrated in Fig.~\ref{fig:overview}.C. As depicted in Fig.~\ref{fig:overview}.F, a prototype was manufactured with an array of 11 LRAs. They were glued under an elongated printed circuit board (PCB) substrate for seamless integration of driving electronics and sensors (see Methods). 

\subsection{Deep sub-wavelength band gap}

\begin{figure}[!t]
\centering
\includegraphics[width=180mm]{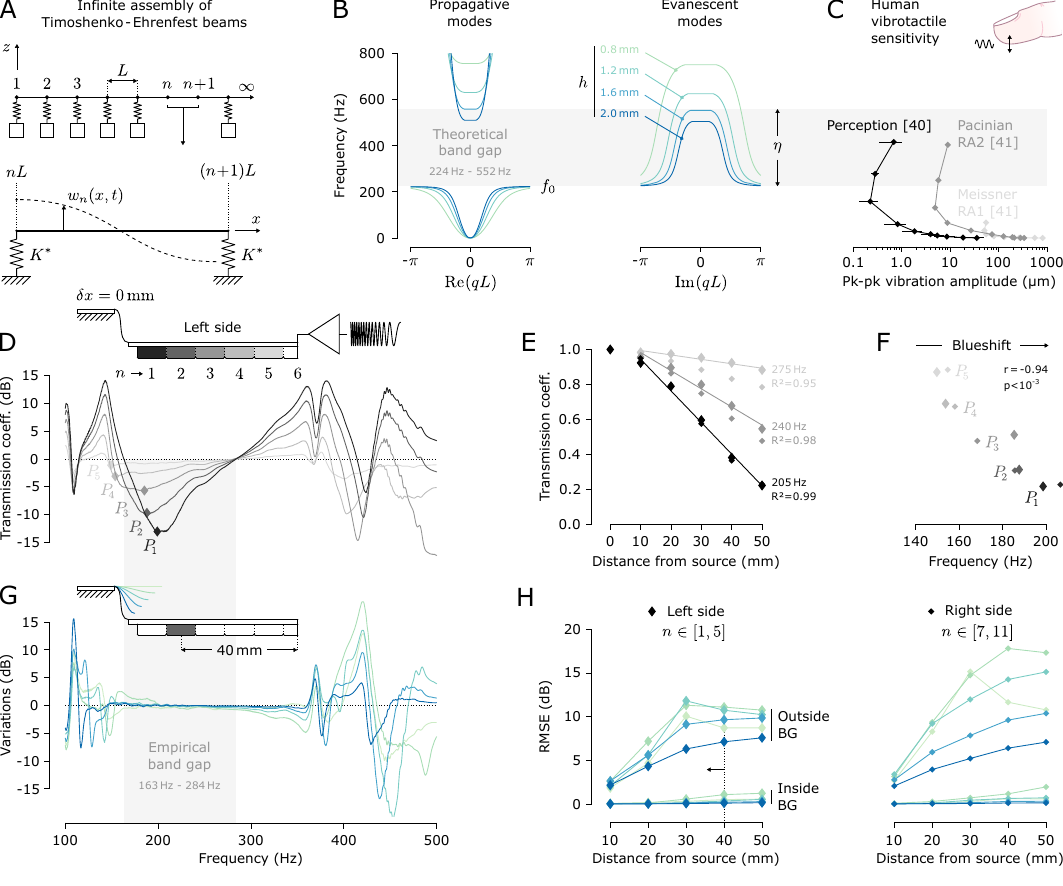}
\caption{\scriptsize{\textbf{Band gap analysis.} \textbf{A.} Metamaterial as an infinite series of Timoshenko-Ehrenfest beams with effective springs, $K^*$\!, attached to each junction, modeling the resonant action of LRAs. \textbf{B.} Dispersion graphs showing complete band gaps for different PCB thicknesses. Real and imaginary parts of the wave vector yield propagative and evanescent modes, respectively. \textbf{C.} Human vibrotactile threshold defined, at a given frequency, as the minimal peak-to-peak vibration amplitude perceivable. Reproduced from~\cite{MountcastleEtAl-72, BolanowskiEtAl-88}. \textbf{D.} Left-hand side transmission coefficient with respect to the central unit cell, $n=6$. \textbf{E.} Transmission coefficient as a function of distance from the vibration source. \textbf{F.} Frequency shift of the attenuation peaks. \textbf{G.} Variations in transmission coefficients for $n=2$, where shades of blue and green represent increasing levels of flexure tension relative to the reference $\delta x=\SI{0}{\milli\meter}$. \textbf{H.} Frequency-averaged variations in transmission coefficients as a function of distance from the vibration source.}}
\label{fig:band_gap}
\end{figure}

Acoustic metamaterials are usually fine-tuned through time-consuming finite element analysis (FEA). Instead, using an array of low-loss resonators provides a closed-form solution, enabling analytical optimization and a deeper understanding of the underlying physics. To unveil the band structure, the PCB substrate was modeled as an infinite series of Timoshenko-Ehrenfest beams of length $L$, as shown in Fig.~\ref{fig:band_gap}.A. Each junction was described by an effective stiffness, $K^*$\!, representing the reaction force of an LRA (see Supplementary Fig.~2.A). By virtue of its periodic symmetry, the problem was reduced to that of a single segment endowed with both continuity and Floquet-Bloch boundary conditions (see Supplementary Information). It yielded the theoretical dispersion diagrams in Fig.~\ref{fig:band_gap}.B, revealing a complete band gap wherein only an evanescent field established. The band gap width, $\eta$, followed an inverse power-law relationship with PCB thickness, $\eta=34.4\,h^{\minus\,0.43}-f_0$, demonstrating excellent agreement ($\mathrm{RMSE}=\SI{1.3}{\hertz}$, see Supplementary Fig.~2.B). In extremely thin substrates, the upper bound of the band gap tends to infinity. However, this would not occur in practical implementations due to high-order modes, unaccounted for in this model, which would otherwise alter the band structure. A \SI{1.6}{\milli\meter} thickness, prevalent for PCBs, was chosen to induce a theoretical band gap from \SI{223}{\hertz} to \SI{552}{\hertz}, aligning with the range of peak vibrotactile sensitivity mediated by Pacinian corpuscles~\cite{MountcastleEtAl-72, BolanowskiEtAl-88}, as shown in Fig.~\ref{fig:band_gap}.C.

These attenuation qualities were verified experimentally by measuring the transmission coefficient, $\mathrm{T}$, between two distinct unit cells (see Methods). Given the symmetry of our prototype, it was computed with respect to the central unit cell, indexed $n=6$, such as $\mathrm{T} = 20\log(\lvert\Gamma_n\rvert/\lvert\Gamma_6\rvert)$, where $\Gamma_n$ is the Fourier transform of the transverse acceleration of the $n$\textsuperscript{th}\! unit cell, with $n\in[1,11]$. Left-hand side transmission coefficients, for $n\in[1,5]$, are presented in Fig.~\ref{fig:band_gap}.D. Right-hand side results are analogous despite minor discrepancies (see Supplementary Fig.~3.A).

A band gap ranging from \SI{163}{\hertz} to \SI{284}{\hertz} was observed. Theory overestimated its boundaries by 37.4\% and 94.4\%, respectively. This may be attributed to several factors including part-to-part variations among LRA batches, mechanical coupling between the steel casings of adjacent LRAs, and overlooked torsional modes that might have further reduced the band gap. A maximum attenuation of \SI{-13.2}{\deci\bel} at \SI{205}{\hertz} was recorded five unit cells away from the vibration source. As shown in Fig.~\ref{fig:band_gap}.D, the peaks, labeled $P_{\!n}$, are shifted towards higher frequencies as more unit cells are involved. This blueshift is linearly correlated to the transmission coefficient (Pearson correlation coefficient $r=-0.94$, $\mathrm{p}<10^{-3}$), as shown in Fig.~\ref{fig:band_gap}.F. In fact, the attenuation gradually established with the number of unit cells recruited. As it better approximates the theoretical infinite model, the attenuation converges towards a maxima for $f \rightarrow f_0$. A convergence error of only 8\% was measured at $P_1$, five unit cells away from the source. Within the band gap, Floquet-Bloch's theorem predicts that vibration amplitude decays exponentially with the number of unit cells (see Supplementary Information). It followed piecewise linear functions instead ($\forall \, f \!\in [205,240,275]\,\SI{}{\hertz}, \, R^2>0.95$), as shown in Fig.~\ref{fig:band_gap}.E. This might be attributed to the small number of unit cells considered. Nonetheless, assuming a smooth decay, the transmission coefficient would still approach zero asymptotically, in line with Floquet-Bloch's predictions. These linear fits failed to intercept the unity transmission coefficient, indicating nonlinearity near the vibration source.

Our acoustic metamaterial was assumed to be primarily subjected to Lamb waves with an asymmetric A0 mode. Their wavelength, $\lambda=\SI{220}{\milli\meter}$ at \SI{205}{\hertz}, was estimated using finite element analysis (see Supplementary Information). In turn, the dimensionless ratio, $\phi=\lambda\,/L=22$, indicates exceptional subwavelength capabilities. This metamaterial can hinder the propagation of elastic waves with wavelengths 22 times larger than a single unit cell. This is a testament to the effectiveness of leveraging prior industrial efforts directed towards the optimization of resonators for mobile applications.

\subsection{Invariance by changes in boundary impedance}

Spurious wave reflections at the extremities of a beam create standing waves and, ipso facto, destructive interference that could mislead band gap analyses. To mitigate these effects, a common strategy in numerical simulations is to terminate edges with perfectly matched layers that provide near-zero reflection~\cite{Berenger-94, LiuTao-97}. However, this solution is not easily implemented in practice. As an alternative, previous transmission measurements were replicated at varying levels of boundary impedance. This was achieved by suspending the metamaterial on flexures, whose stiffness was adjusted via a tensioning motion $\delta x$ (see Methods, Supplementary Fig.~4, and Movie~1).

Variations in transmission coefficients are presented in Fig.~\ref{fig:band_gap}.G, at five levels of boundary impedance, each obtained by \SI{2}{\milli\meter} increments from the reference signal taken at $\delta x=\SI{0}{\milli\meter}$. Results are given for $n=2$ (see Supplementary Fig.~3.B for remaining values of $n$). The band gap is the only frequency range that remained mostly unaffected by changes in boundary conditions. As shown in Fig.~\ref{fig:band_gap}.H, the RMSE with respect to the reference, $\delta x=\SI{0}{\milli\meter}$, is one to two orders of magnitude lower within the band gap than outside of it. Furthermore, the RMSE outside the band gap increases with the distance from the vibration source, showing significant edge effects otherwise barely seen within the band gap. This provides further evidences that the extraordinary attenuation qualities are intrinsic to the resonant acoustic metamaterial rather than an artifact of suitably designed boundaries.

\subsection{Spatially localized acceleration field}

\begin{figure}[!t]
\centering
\includegraphics[width=180mm]{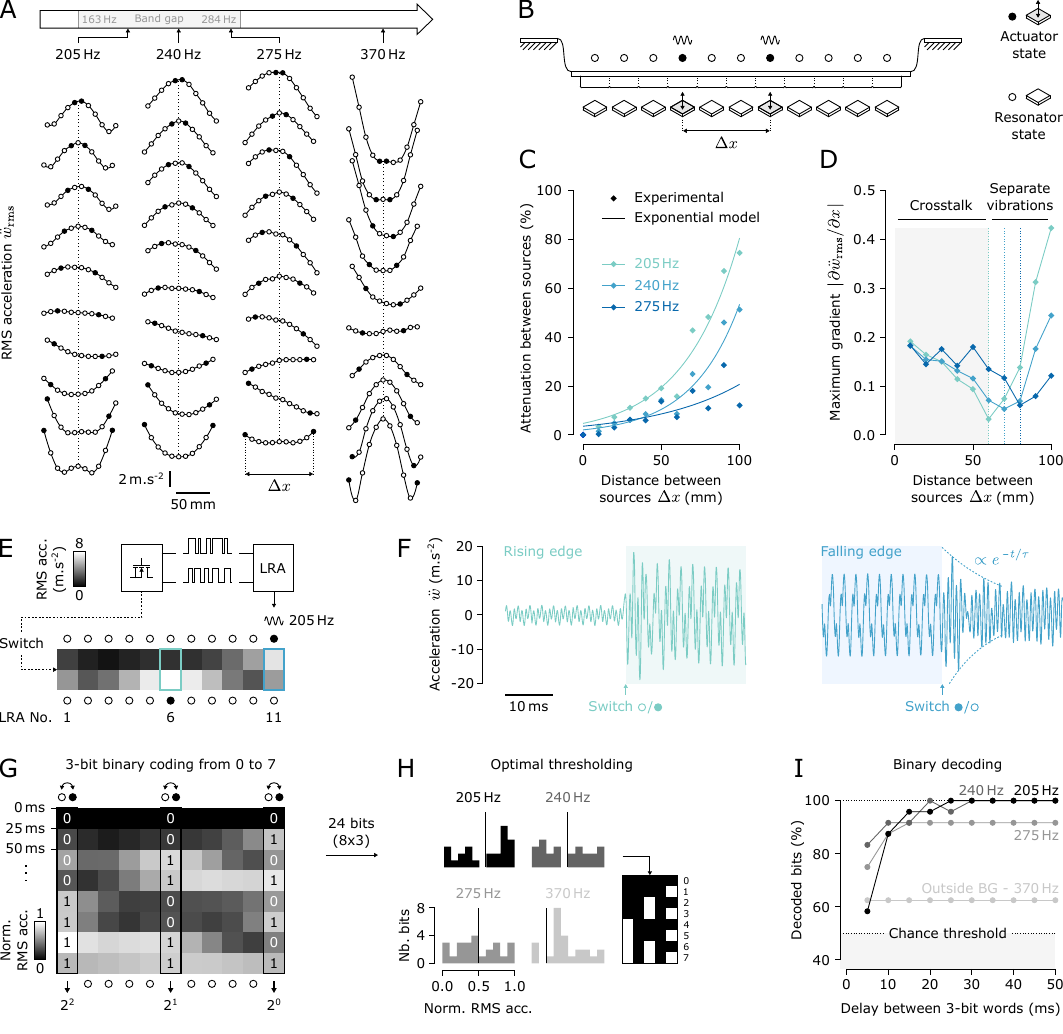}
\caption{\scriptsize{\textbf{High-speed reconfiguration.} \textbf{A.} Interpolated RMS acceleration field for two unit cells actuated. \textbf{B.} Schematic of the experiment with two unit cells driven by sine waves. \textbf{C.} Exponential increase in attenuation with the distance between sources. \textbf{D.} Maximum gradient of the acceleration field. The inflection point marks the transition from crosstalk between both sources to distinct vibration spots. \textbf{E.} Steady-state RMS acceleration with actuation of LRA $n=11$, then commuted to LRA $n=6$, for a \SI{205}{\hertz} carrier. \textbf{F.} Rising and falling edges recorded during the electronic commutation of LRAs $n=6$ and $n=11$, respectively. The dotted line represents the exponential envelope of the LRA settling. \textbf{G.} Steady-state RMS acceleration for 3-bit binary numbers counted from zero to seven in \SI{25}{\milli\second} intervals. \textbf{H.} Decoding sequence with binarization using thresholds optimized for each carrier. \textbf{I.} Decoding success rate as a function of the time delay between each word.}}
\label{fig:reconfiguration}
\end{figure}

To maximize energy efficiency and produce vivid stimuli, LRAs set electronically in an actuator state must be driven with a carrier close to resonance. In turn, the carrier frequency naturally falls within the band gap of a metamaterial made from remaining LRAs in a resonator state. It is, in that sense, self-tuned. By strategically arranging unit cells in both states, vibrations can be confined in localized areas. However, sufficient spacing must be introduced between activated unit cells to minimize crosstalk.

This was investigated by driving two unit cells with a sinusoidal carrier and increasing their separation, $\Delta x$, in \SI{10}{\milli\meter} increments up to \SI{100}{\milli\meter}, as illustrated in Fig.~\ref{fig:reconfiguration}.B. The RMS acceleration profiles along the metamaterial were interpolated by cubic splines, as shown in Fig.~\ref{fig:reconfiguration}.A. Within the band gap, adjacent sources induced a unique vibration spot. To create distinct spots, they had to be separated by at least \SI{60}{\milli\meter}. This is evident from the inflection point in the acceleration gradient, as shown in Fig.~\ref{fig:reconfiguration}.D. To achieve the sharpest vibration spots, characterized by the highest gradient, $\Delta x$ should be maximized. In practise, this was limited by wave reflections when LRAs were activated close to the extremities. This effect was particularly pronounced beyond the band gap, at \SI{370}{\hertz}, where destructive interference led to a minimal acceleration, even near vibration sources.
Further insights into the attenuation qualities were gained by analyzing a larger number of unit cells than in the previous experiment. This corroborates the exponential decay of evanescent waves predicted by Floquet-Bloch theory ($\forall \, f \!\in [205,240,275]\,\SI{}{\hertz}, \, \mathrm{RMSE}\leq 5.9\%$), as depicted in Fig.~\ref{fig:reconfiguration}.C.

\subsection{High-speed reconfiguration}

Our active acoustic metamaterial was specifically designed for dynamic reconfiguration. This was exemplified by electronically switching the powered unit cell from the distal one to the central one, corresponding to $n=11$ and $n=6$, respectively, as illustrated in Fig.~\ref{fig:reconfiguration}.E. The transient acceleration during both rising and falling edges is shown in Fig.~\ref{fig:reconfiguration}.F. The falling edge was followed by an exponential decay with a time constant $\tau = \SI{22.2}{\milli\second}$, slightly overestimated by 28.4\% by the second-order model, $1/\zeta\,\omega_0 = \SI{28.5}{\milli\second}$. The rising edge established even faster due to the forced response. These results support the use of electromagnetic actuation in combination with MOSFET-based commutation to enable swift reconfiguration. In fact, the limiting factor is the lack of damping, rather than electronics. However, increasing damping could inhibit the resonant band gap.

The ability to rapidly shape vibration patterns holds potential for information storage and transmission, which can be quantified by the bandwidth. It was explored by encoding binary words spatially in the wave field. Limiting them to 3 bits, $n\in\{1,6,11\}$, ensured they were sufficiently spaced apart with $\Delta x=\SI{50}{\milli\meter}$. Although slightly less than the previously defined threshold, it effectively reduced crosstalk between each bit while maintaining compactness. The steady-state RMS acceleration for 3-bit binary numbers counted from zero to seven at \SI{25}{\milli\second} intervals is given in Fig.~\ref{fig:reconfiguration}.G (see Supplementary Movie~2). These patterns can be simply deciphered by thresholding the histogram of RMS acceleration. For each carrier, an optimal threshold can be found numerically, as shown in Fig.~\ref{fig:reconfiguration}.H. The decoding success rates are given in Fig.~\ref{fig:reconfiguration}.I. Outside the band gap, it remained close to chance level, independent of the delay, as uncontrolled propagation blurred the binary words. Nonetheless, an exact retrieval is possible for carriers within the band gap, provided that a minimum delay of \SI{25}{\milli\second} is respected between each word. This confirms previous observations on the time required to settle a logic state. Despite using electromagnetic unit cells, our metamaterial achieved a reconfiguration time comparable to that of piezoelectric metalenses (\SI{13}{\milli\second})~\cite{PopaEtAl-15}. As a result, it can emulate a lossless vibratory display with a $\SI[inter-unit-product = \!\cdot\!]{120}{\bit\per\second}$\! rate at a \SI{50}{\milli\meter} spatial resolution.

\subsection{Application to localized vibrotactile feedback}

\begin{figure}[!b]
\centering
\includegraphics[width=88mm]{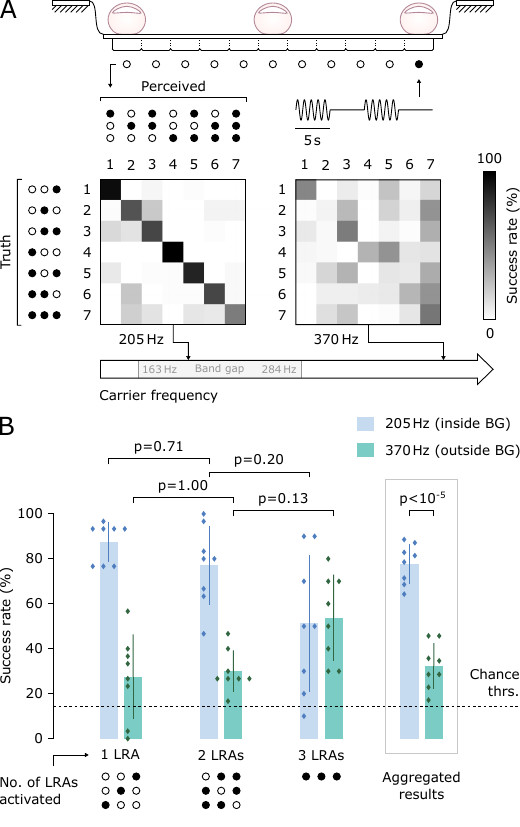}
\captionsetup{width=88mm}
\caption{\scriptsize{\textbf{Perceptual experiment.} \textbf{A.} Confusion matrices from the perception of localized vibration patterns that encode binary representations of integers from one to seven. Results averaged across all participants. \textbf{B.} Average success rate of the matching task, either aggregated or categorized by the number of LRAs activated simultaneously. Error bars represent one standard deviation. The dashed line represents a chance threshold of 1/7.}}
\label{fig:perception}
\end{figure}

To evaluate the potential to convey tactile information, we devised a perceptual matching task (see Methods). Participants were asked to distinguish between seven different vibration patterns, corresponding to the binary conversion of integers from one to seven. Confusion matrices averaged across participants are given in Fig.~\ref{fig:perception}.A for two sinusoidal carriers, \SI{205}{\hertz} and \SI{370}{\hertz}, within and beyond the band gap, respectively. A two-way repeated measures ANOVA revealed significant effects of both the vibration pattern ($\mathrm{F}(6,98)=4.6,\,\mathrm{p}<10^{-3}$) and the carrier frequency ($\mathrm{F}(1,98)=119.9,\,\mathrm{p}<10^{-5}$). This strongly indicates that the perception of local stimuli was mediated by the band gap. These findings are consistent with the acceleration fields depicted in Fig.~\ref{fig:reconfiguration}.G. The interaction term between pattern and frequency significantly impacted the success rate ($\mathrm{F}(6,98)=6.5,\,\mathrm{p}<10^{-5}$). To further explore this effect, the data was segregated based on the number of simultaneously activated LRAs and a two-way repeated measures ANOVA was conducted. A post-hoc Tukey-Kramer test revealed that an activation of one or two LRAs yielded similar success rates, both in ($\mathrm{p}=0.71$) and out ($\mathrm{p}=1.00$) of the band gap, as shown in Fig.~\ref{fig:perception}.B. Despite the lack of statistical significance ($\mathrm{p}=0.13$ and $\mathrm{p}=0.20$), a notable contrast was observed when all three LRAs were activated. In that case, the band gap had a negligible effect on perception as participants seemed to confuse multiple vibration points with vibration spread. The increased variance in success rate illustrates this confusion. The mislocalization of tactile stimuli may have also been caused by a funneling illusion, as described by von Békésy~\cite{vonBekesy-59, ChenEtAl-03}.

Overall, participants could accurately perceive tactile messages spatially encoded on three bits with a success rate of $81\%$. Interestingly, the viscoelastic skin tissues in contact with the acoustic metamaterial did not alter its operational principles, as it still greatly enhanced the perception of local stimuli. However, for frequencies beyond its operating range, the success rate dropped to a mere $34\%$, highlighting the inherent limitations in rendering localized haptics on conventional continuous media.

\subsection{Wavefield control along a pre-defined path}

\begin{figure}[!b]
\centering
\includegraphics[width=180mm]{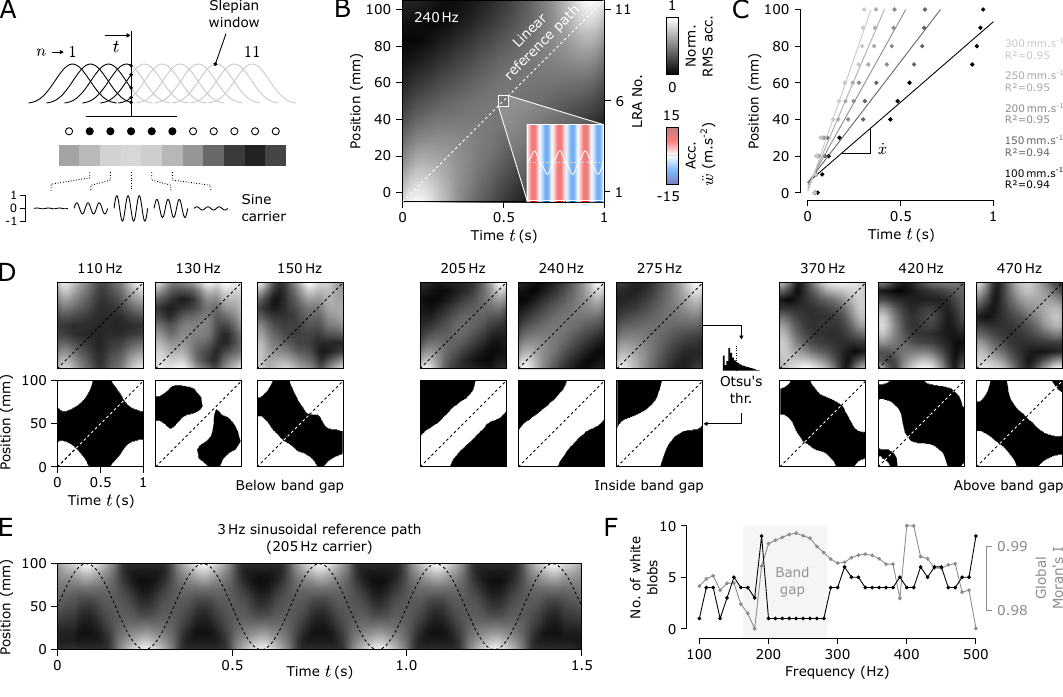}
\caption{\scriptsize{\textbf{Wavefield guiding along spatiotemporal paths.} \textbf{A.} Activation sequence of LRAs supplied with Slepian-windowed carrier signals. Up to five LRAs were activated simultaneously. \textbf{B.} Recorded acceleration field following a linear spatiotemporal reference path represented by the dashed line. \textbf{C.} Evolution of the vibration spot for various path speeds, and corresponding linear fits. \textbf{D.} Spatiotemporal vibration maps. The top row was obtained by bi-linear interpolation of the RMS acceleration, and the bottom row by Otsu's binarization. \textbf{E.} Sinusoidal reference path sweeping the metamaterial lengthwise at \SI{3}{\hertz}. \textbf{F.} Metrics for assessing path-following effectiveness.}}
\label{fig:waveguide}
\end{figure}

Our metamaterial goes beyond discrete reconfiguration through 3-bit binary words. It supports complex transitions, such as the continuous movement of a vibration spot along a desired path. To that end, the proposed method involved driving unit cells with overlapping windowed signals. Specifically, a sinusoidal carrier, windowed by a zero-order Slepian sequence was employed. It minimized spectral leakage, limiting high-frequency artifacts that might have otherwise extended beyond the band gap (see Supplementary Fig.~5). Overlapping adjacent signals by 80\% was found empirically to ensure continuous traveling of the acceleration field. This resulted in the activation of up to five unit cells simultaneously, as shown in Fig.~\ref{fig:waveguide}.A.

An experiment was conducted aiming at continuously guiding a vibration spot along a linear reference path. The Slepian sequence was timed accordingly. The RMS acceleration field was interpolated to compensate for the limited spatial resolution of the sensor array. This resulted in the spatiotemporal vibration maps depicted in Fig.~\ref{fig:waveguide}.B and Fig.~\ref{fig:waveguide}.D, which were further processed as images. To reveal regions with significant differences in vibration levels, they were binarized using Otsu's thresholding method. Only the central panel, for carrier frequencies inside the band gap, features a unique vibration spot, shown in white, centered around the reference path. Such single anti-node moving both in time and space demonstrates the effectiveness of our approach. This is further supported by the global Moran's index, which measures spatial autocorrelation. Clustering, typical of an isolated vibration spot, indeed led to the highest autocorrelation values, as evident in Fig.~\ref{fig:waveguide}.F. The reference path was traveled at velocities, $\dot{x}$, up to $\SI[inter-unit-product=\!\cdot\!]{300}{\milli\meter\per\second}$ (see Supplementary Movie~3). As shown in Fig.~\ref{fig:waveguide}.C for a \SI{205}{\hertz} carrier, successful path following was achieved regardless of velocity ($\forall \, \dot{x} \in \{100,150,\dots,300\}\,\SI[inter-unit-product=\!\cdot\!]{}{\milli\meter\per\second}\!, \, R^2>0.94$). This capacity extends to sinusoidal paths, as shown in Fig.~\ref{fig:waveguide}.E (see Supplementary Movie~4). Therefore, the proposed metamaterial enables continuous steering of low-frequency vibration spots in both time and space. This holds great promise for rendering compelling sensations of tactual apparent motion~\cite{Burtt-17}.

\section{Discussion}

This study tackled the issue of invariant acoustic properties in passive structures by introducing a new type of active metamaterial. It consists of a series of dual-state unit cells, either resonators when left disconnected or vibration sources when powered. Conveniently, these unit cells can be made from off-the-shelf LRAs, commonly found in smartphones. They create a self-tuned metamaterial with a band gap that intrinsically includes the optimal actuation carrier frequency. A major strength of this approach is its potential to create acoustic metamaterials with minimal knowledge, serving as a turnkey solution for haptic designers.

Comprehensive theoretical and experimental investigation of a 1D prototype unveiled a band gap ranging from about \SI{160}{\hertz} to \SI{290}{\hertz}, with a maximum attenuation of \SI{13.2}{\deci\bel} at \SI{205}{\hertz}, after five unit cells. This effect was conclusively attributed to local resonances, rather than boundary conditions. A bespoke controller supports real-time reconfiguration of the metamaterial in less than \SI{25}{\milli\second}. This enables dynamic shaping of vibration patterns with a bandwidth of about $\SI[inter-unit-product = \!\cdot\!]{120}{\bit\per\second}$\!. In addition, these unit cells exhibit deep subwavelength characteristics by stopping the propagation of waves with wavelengths 22 times their size. This paves the way for materials that mediate vibrotactile feedback in a compact form factor.

A perceptual experiment demonstrated how human subjects were successfully able to retrieve 3-bit messages ingrained spatially on a \SI{110}{\milli\meter}-long prototype. Its ability to steer a vibration spot along a smooth spatiotemporal path also offers a promising solution for creating compelling illusions of apparent motions. In turn, this work has direct translational applications as a live communication device for the visually impaired as it already follows the Perkins Brailler layout. Beyond manifest tactile applications in multi-touch refreshable displays, storing bits of elastic energy spatially with a dynamic allocation might be an interesting use case for purely analog computing solutions.

\newpage
\section{Methods}

\subsection{Prototype}

An acoustic metamaterial was manufactured with an array of 11 LRAs (VLV101040A, Vybronics). They were glued on an elongated PCB with a thin layer of cyanoacrylate adhesive (454, Loctite), resulting in a rigid bond that effectively transmitted vibrations. Both extremities of the metamaterial were suspended by flexures cut in a \SI{100}{\micro\meter}-thick polyimide film (Kapton, DuPont). A pop-up structure was designed within this rigid-flex assembly to provide adjustable tensioning of the flexures via a linear motion $\delta x$ (see Supplementary Fig.~4). Manual setting is demonstrated in Supplementary Movie~1. This allowed the study of dispersive effects due to changes in boundary stiffness, which increased monotonically with flexure tension. These flexures also created efficient routing paths for electrical signals ($\sim 40$ traces), avoiding undesired stiffness from cables. The device was mounted on a \SI{1.3}{\kilo\gram} brass block fitted with rubber feet that filtered out external disturbances. The prototype is illustrated in Fig.~\ref{fig:overview}.F.

\subsection{Vibration generation}

The embedded control unit is a 32-bit microcontroller board (Teensy 4.0, PRJC) running at \SI{1.0}{\giga\hertz}. The LRAs were driven in open-loop through H-bridge MOSFET drivers (DRV8837, Texas Instruments), fed by pulse width modulated (PWM) signals coded on 10 bits. To meet the stringent timing requirements, output waveforms were generated offline and stored in look-up tables. The entire device was powered through USB with \SI{5}{\volt}, lowered down to \SI{2.5}{\volt} with a buck converter (Okami OKL, Murata) in order to supply the LRAs. A large bank of decoupling capacitors covered the transient current intakes from the actuators. A simplified control scheme and the PCB layout are given in Supplementary Fig.~6. For the characterization of the band gap, the central unit cell was set to an actuator state and supplied with a linear chirp from DC to \SI{500}{\hertz}, while the remaining cells were left in a resonator state. To avoid any undesired distortions that may be caused by the digital signal generation, a 16-bit analog signal output from an acquisition card (USB-6343, National Instruments) was used instead, but only for the purpose of this analysis. The signal was then fed to a class-D amplifier (TPA3112D1, Texas Instruments).

\subsection{Full-field vibrometry}

To provide real-time, full-field vibrometry, the prototype was fitted with 11 analog accelerometers (ADXL335, Analog Devices), each centered to its corresponding LRA. Only their $z$-axes were connected, followed by first-order low-pass RC filters with a \SI{500}{\hertz} cut-off ($f_c=1/2\pi RC$). The acceleration field along the beam was recorded at a sensitivity of $\SI{300}{\milli\volt}\!\cdot\!\mathrm{g^{-1}}$\! and sampled with a 12-bit analog-to-digital converter (ADC) at \SI{10}{\kilo\hertz}. A maximum delay of only \SI{4}{\micro\second} was measured between successive analog input readings, thus providing the excellent synchronicity required for the transient analysis of active reconfigurations. To maintain synchronicity during the experiments, data were stored on a \SI{1}{\mega\byte} RAM buffer rather than directly exported through USB.

\subsection{Impulse response of a unit cell}

A \SI{1}{\milli\second} square pulse of $2.28\pm\SI{0.03}{\volt}$ was fed to an LRA through a MOSFET, and repeated 50 times. This circuit was designed to disconnect the coil immediately after pulsing, which prevented the occurrence of undesirable eddy current damping. The LRA was mechanically grounded. Its casing was removed to provide access for the beam of a \SI{632.8}{\nano\meter} HeNe laser interferometer (LSV-2500-NG, SIOS) pointing towards the moving mass, as illustrated in Fig.~\ref{fig:overview}.E. Data were digitized on 16 bits at a sampling rate of \SI{100}{\kilo\hertz} by an acquisition card (PCI-6121, National Instruments), followed by a zero-lag, two-pole Butterworth low-pass filter with a \SI{1}{\kilo\hertz} cut-off. It resulted in a noise floor of \SI{18}{\nano\meter} RMS, sufficient to resolve minute vibrations.

\subsection{Participants and protocol}

Eight volunteers (5 males, 3 females) aged 25.1 $\pm$ 2.8 (mean $\pm$ std) participated in this experiment. The study was conducted with the approval of Sorbonne University Ethics Committee (CER-2021-078) and the participants gave their written informed consent. They were instructed to place their fingertips on the acoustic metamaterial at the three specific locations indicated in Fig.~\ref{fig:perception}.A. Their fingers were covered and they wore noise-cancelling headphones to prevent any visual or auditory cues. Flexures were tensioned to $\delta x=\SI{10}{\milli\meter}$. The experiment consisted in a matching task in which participants had to recognize localized vibrations patterns, corresponding to the binary conversion of integers from one to seven. Each spatial pattern was randomly repeated 20 times and was displayed with sinusoidal signals at either \SI{205}{\hertz} or \SI{370}{\hertz}. Each tactile stimulus lasted \SI{5}{\second} and the participants had to give an answer within the next \SI{5}{\second}. The vibration amplitude was kept constant. A total of 140 trials were presented in 2 blocks, separated by a \SI{5}{\minute} break.

\backmatter

\section*{Data availability}

The data that support the findings of this study are available from the corresponding author upon reasonable request.

\section*{Code availability}

The codes that support the findings of this study are available from the corresponding author upon reasonable request.

\section*{Acknowledgement}

This work was supported by the French National Research Agency (ANR-20-CE33-0013 Maptics).

\section*{Author contributions}

T.D. conceived the research, designed and manufactured the metamaterial, conducted analytical and experimental studies, analyzed the data, and wrote the original draft of the manuscript. D.G., S.H., and V.H. secured fundings and supervised the work. All authors reviewed the manuscript.

\section*{Additional information}

\bmhead{Supplementary information}

The online version contains supplementary material available at https://doi.org/xx.xxxx/xxx.

\bmhead{Competing interests}

The authors declare no competing interests.

\bmhead{Reprints and permission}

Reprints and permission information is available online at http://npg.nature.com/
reprintsandpermissions/

\newpage
\bibliography{2024_TD_SH_DG_VH_main}

\end{document}


\title[Article Title]{\centering Supplementary Information}

\author*[]{\fnm{Thomas} \sur{Daunizeau$^{\ast}$}}\email{\href{mailto:thomas.daunizeau@sorbonne-universite.fr}{thomas.daunizeau@sorbonne-universite.fr}}
\author[]{\fnm{Sinan} \sur{Haliyo}}
\author[]{\fnm{David} \sur{Gueorguiev}}
\equalcont{}
\author[]{\& \fnm{Vincent} \sur{Hayward$^{\S}$}}
\equalcont{These authors contributed equally to this work.\quad $^{\S}$Deceased.\phantom{\hspace{29.95mm}}}
\affil[]{\orgdiv{Sorbonne Université, CNRS, ISIR}, \postcode{F-75005} \city{Paris}, \country{France}}

\maketitle

\newpage
\section{Dynamic analysis of an LRA}

\subsection{Investigation of eddy current damping}

The dynamic response of an LRA was evaluated by measuring its impulse response. It was driven by \SI{1}{\milli\second} square pulses of $2.28\pm\SI{0.03}{\volt}$, repeated 50 times. A typical pulse train is illustrated in Supplementary Fig.~\ref{fig:supplementary_impulse_response}.A. Following the falling edges, two configurations can be achieved using the MOSFET driver: leaving the coil in open circuit or shorting it. As shown in Fig.~1.B, the first configuration, with $i=0$, establishes a low-loss resonator with a high quality factor, $Q=19.9$. In contrast, the second configurations shown in Fig.~\ref{fig:supplementary_impulse_response}.B introduces magnetic damping through eddy currents, $i \neq 0$, potentially offering a flat frequency response capable of seamless vibration propagation. Its response, averaged over 50 trials, is given in Supplementary Fig.~\ref{fig:supplementary_impulse_response}.C. It reveals a slightly lower quality factor, $Q_d=17.8$ (-10.5\%), indicating enhanced damping. This is further supported by the difference in gain, $|H|-|H_d|$, shown in Supplementary Fig.~\ref{fig:supplementary_impulse_response}.D. However, this additional damping was insufficient to completely suppress the resonance. This would require closed-loop control, which is beyond the scope of this study.

\begin{figure}[!h]
\centering
\includegraphics[width=88mm]{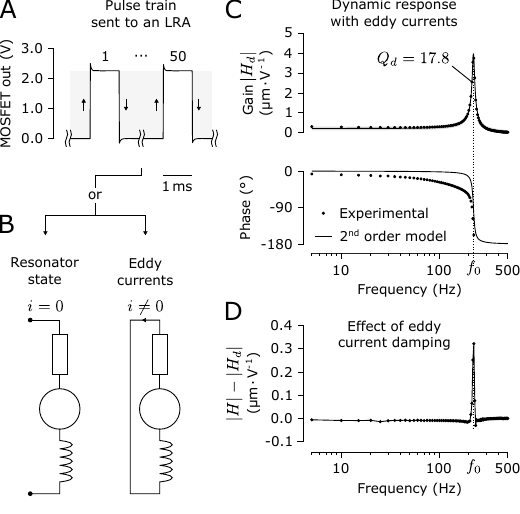}
\captionsetup{width=88mm}
\caption{\scriptsize{\textbf{Detailed study of LRA dynamics.} \textbf{A.} Typical pulse train sent to an LRA to evaluate its dynamic response. \textbf{B.} Equivalent wiring of the two possible configurations after a falling edge. \textbf{C.} Impulse response of an LRA subjected to eddy currents, averaged over 50 trials. \textbf{D.} Gain reduction due to eddy current damping, centered around the resonance.}}
\label{fig:supplementary_impulse_response}
\end{figure}

\subsection{Second-order LTI models}

The impulse responses with and without eddy currents, shown in Supplementary Fig.~\ref{fig:supplementary_impulse_response}.B and Fig.~1.E, respectively, were averaged over 50 trials and fitted with second-order linear time-invariant (LTI) models using nonlinear least squares. The transfer function $H$ between the output displacement $z(t)$ and the input voltage $v(t)$ is given by,
\begin{equation}\label{eq:second_order_LRA}
H(s) = \frac{Z(s)}{V(s)} = \frac{G\,\omega_0^2}{s^2 + 2\,\zeta\omega_0 \, s+ \omega_0^2}
\, ,
\end{equation}
where $s$ is the Laplace variable, $G$ is the DC gain, $\omega_0$ is the undamped natural frequency, and $\zeta$ is the damping ratio. These parameters were suffixed with the subscript $d$ to describe the state with eddy current damping.

\section{Metamaterial model and characterization}

\subsection{Infinite series of Timoshenko-Ehrenfest beams}

The elongated PCB substrate was modeled as a beam made from a glass-epoxy laminate, commercially known as FR-4. Its multiple layers of finely woven glass fiber cloth were hypothesized to confer quasi-isotropic and homogeneous flexural properties, along with linear elasticity. As previously shown, an LRA can be approximated by a lumped mass-spring model with negligible damping. Potential contributions from its metal casing were discarded. As depicted in Fig.~2.A, this acoustic metamaterial can be divided into beam segments following the spatial periodicity, $L$, along the $x$-axis. Lumped LRA models were inserted at each junction. By virtue of its periodic symmetry, this problem was reduced to that of a single segment endowed with boundary conditions. The $n$\textsuperscript{th} segment, $\forall x \in [nL,(n+1)L]$, is a thick beam whose transverse displacement, $w_n$, is described by the Timoshenko-Ehrenfest theory, i.e. taking into account both rotary inertia and shear deformation. It was thus to satisfy the following flexural wave equation,
\begin{equation}\label{eq:wave_timoshenko_LRA}
EI \pdv[4]{w_n}{x} + \rho S \pdv[2]{w_n}{t} - \rho I \Big(1+\frac{E}{\kappa G} \Big) \pdm 4{w_n} 2x 2t + \frac{{\rho}^2 I}{\kappa G}\pdv[4]{w_n}{t} = 0
\, ,
\end{equation}
where $\rho$ is the density, $E$ is Young's modulus, $G = E/2(1+\nu)$ is the shear modulus, $\nu$ is Poisson's ratio, $\kappa \approx 5/6$ is the shear coefficient of a rectangular cross section, $S = bh$ is the cross-sectional area, $I = bh^3\!/12$ is the moment of inertia, $b$~is the width, and $h$ is the thickness. The separation of variables provides a general solution decomposed in spatial and temporal components as in $w_n(x,t) = A_n(x)\, e^{\,j(\omega t+\phi)}$. Therefore, equation~\eqref{eq:wave_timoshenko_LRA} becomes a fourth order differential equation,
\begin{equation}\label{eq:wave_reduced_timoshenko_LRA}
\dv[4]{A_n}{x} - \alpha \dv[2]{A_n}{x} - \beta A_n = 0
\, ,
\end{equation}
with,
\begin{equation}\label{eq:alpha_beta}
\alpha = -\rho\, {\omega}^2 \Big( \frac{1}{E}+\frac{1}{\kappa G} \Big)\, , \quad \textrm{and} \quad \beta = \frac{\rho\, {\omega}^2}{E} \Big( \frac{S}{I}-\frac{\rho\, {\omega}^2}{\kappa G} \Big)
\, .
\end{equation}\\
The spatial amplitude, $A_n$, solution of equation~\eqref{eq:wave_reduced_timoshenko_LRA} is a combination of propagating and evanescent waves $A_{i,n}$, such as,
\begin{equation}\label{eq:modal_timoshenko}
A_n(x) = \sum_{i=1}^{4} A_{i,n}(x) = \sum_{i=1}^{4} C_{i,n}\, k_i^{-3} e^{\,k_i(x-nL)}
\, ,
\end{equation}
where $C_{i,n}$ are complex constants dependent on the boundary conditions and $k_i$ are complex wavenumbers defined by the following dispersion relation,
\begin{equation}\label{eq:complex_wave_numbers}
k_i = \frac{{(-1)}^{\lfloor i/2 \rfloor}}{\sqrt{2}}\sqrt{\alpha + {(-1)}^i \sqrt{{\alpha}^2+4\beta}} \quad \textrm{for} \quad i \in [1,2,3,4] 
\, ,
\end{equation}
where ${\lfloor \cdot \rfloor}$ is the floor function. Additional analytical information is available in~\cite{Diaz-de-AndaEtAl-05, YuEtAl-06}.

\subsection{Continuity conditions}

Boundary conditions must be applied to factor in the contribution of LRAs and connect adjacent segments. Ensuring the continuity of transverse displacement, slope, bending moment, and shear force at the interface between each segment yields the following respective conditions at any junction $n$,
\begin{align}
    \phantom{aa}
    &\begin{aligned}
    \mathllap{\evalat{A_n}{x=nL}} \, &= \, \mathrlap{\evalat{A_{n-1}}{x=nL}}
    \end{aligned}
    \\[2mm]
    &\begin{aligned}
    \mathllap{\evalat[\Big]{\dv{A_n}{x}}{x=nL}} \, &= \, \mathrlap{\evalat[\Big]{\dv{A_{n-1}}{x}}{x=nL}}
    \end{aligned}
    \\[2mm]
    &\begin{aligned}
    \mathllap{EI \evalat[\Big]{\dv[2]{A_n}{x}}{x=nL}} \, &= \, \mathrlap{EI \evalat[\Big]{\dv[2]{A_{n-1}}{x}}{x=nL}}
    \end{aligned}
    \\[2mm]
    &\begin{aligned}
    \mathllap{EI \evalat[\Big]{\dv[3]{A_n}{x}}{x=nL} \! + K^* \evalat{A_n}{x=nL}} \, &= \, \mathrlap{EI \evalat[\Big]{\dv[3]{A_{n-1}}{x}}{x=nL}}
    \end{aligned}
\end{align}
where $K^*$ is a variable dimensionally equivalent to a stiffness, derived from the dynamic equilibrium of forces along the $z$-axis, and defined by, $\forall \, \omega\neq \omega_0$,
\begin{equation}\label{eq:shear_force}
K^* = \frac{m}{1/{\omega}^2-1/{\omega_0}^2}
\, ,
\end{equation}
\\
where $m$ is the lumped mass and $\omega_0$ is the resonance frequency of a LRA, defined in equation~\eqref{eq:second_order_LRA}. The effective stiffness, $K^*$\!, transcribes the action of LRAs and illustrates how local resonances contribute to the creation of a band gap. As shown in Supplementary Fig.~\ref{fig:supplementary_theory}.A, $K^*\!< 0$ when $\omega > \omega_0$, which yields a shear force opposed to motion, at the origin of the extraordinary attenuation properties of acoustic metamaterials. In high frequencies, the effective stiffness tends towards $\minus K$, which is equivalent to a beam laying on a bed of springs. Conversely, $K^*\!> 0$ when $\omega < \omega_0$, which tends to amplify transverse vibrations. The negative effective stiffness, $K^*$\!, should not be misconstrued as an indication of a negative modulus in the acoustic metamaterial. Rather, it is a mere algebraic simplification devoid of any physical attributes. It is pertinent to highlight that this resonant acoustic metamaterial is, in fact, a canonical instance of an effective negative mass density structure.
\\
\\
During intended use of tactile applications, a skin patch will be in contact with the metamaterial, which will increase the effective stiffness $K^*$\!. The elastic constants of epidermal tissues reported in the literature are an order of magnitude smaller than $K^*$\!. While skin elasticity may be negligible, the effect of viscoelastic damping, prominent above \SI{100}{\hertz}~\cite{WiertlewskiHayward-12}, was assessed experimentally. As shown in Fig.~4, results from the perceptual experiment indicate that tactile perception aligns with the metamaterial band gap, confirming that the impact of skin is minimal.
\begin{figure}[!h]
\centering
\includegraphics[width=88mm]{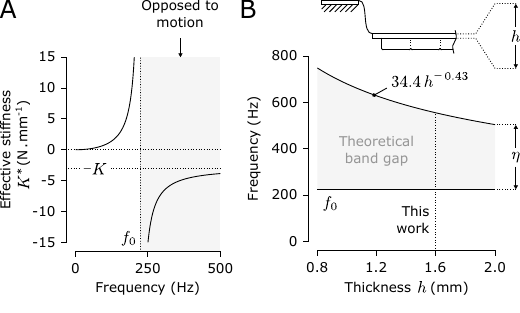}
\captionsetup{width=88mm}
\caption{\scriptsize{\textbf{Theoretical analysis.} \textbf{A.} Effective stiffness as a function of frequency, where negative values result in a shear force proportional and opposite to the direction of motion. \textbf{B.} Band gap boundaries as a function of the FR-4 substrate thickness.}}
\label{fig:supplementary_theory}
\end{figure}
\\
The boundary conditions previously listed can be factorized in matrix form as,
\begin{equation}\label{eq:matrix_boundary}
\mathbf{C_n} = \mathbf{X^{-1}} \mathbf{Y}\, \mathbf{C_{n-1}}
\, ,
\end{equation}
\begin{equation}\label{eq:matrix_X}
\mathbf{X} =
\begingroup
\setlength\arraycolsep{6pt}
\begin{bmatrix} 
k_1^{-3} & k_2^{-3} & k_3^{-3} & k_4^{-3}  \\
k_1^{-2} & k_2^{-2} & k_3^{-2} & k_4^{-2}  \\
k_1^{-1} & k_2^{-1} & k_3^{-1} & k_4^{-1}  \\
1 \! + \! k_1^{-3} K^*\!/EI & 1 \! + \! k_2^{-3} K^*\!/EI & 1 \! + \! k_3^{-3} K^*\!/EI & 1 \! + \! k_4^{-3} K^*\!/EI  \\
\end{bmatrix}
\endgroup
\, ,
\end{equation}
\\
\begin{equation}\label{eq:matrix_Y}
\mathbf{Y} =
\begingroup
\setlength\arraycolsep{10pt}
\begin{bmatrix} 
k_1^{-3}e^{k_1 L} & k_2^{-3}e^{k_2 L} & k_3^{-3}e^{k_3 L} & k_4^{-3}e^{k_4 L}  \\
k_1^{-2}e^{k_1 L} & k_2^{-2}e^{k_2 L} & k_3^{-2}e^{k_3 L} & k_4^{-2}e^{k_4 L}  \\
k_1^{-1}e^{k_1 L} & k_2^{-1}e^{k_2 L} & k_3^{-1}e^{k_3 L} & k_4^{-1}e^{k_4 L}  \\
e^{k_1 L} & e^{k_2 L} & e^{k_3 L} & e^{k_4 L}  \\
\end{bmatrix}
\endgroup
\, , \quad
\mathbf{C_n} = 
\begin{bmatrix} 
C_{1,n} \\
C_{2,n} \\
C_{3,n} \\
C_{4,n} \\
\end{bmatrix}
\, .
\end{equation}

\subsection{Floquet-Bloch theorem}

To reveal the band structure, the metamaterial was assumed to be of infinite periodicity along the $x$-axis, i.e. $n\rightarrow\infty$. Accordingly, the Floquet-Bloch's theorem states that the solution of the wave equation~\eqref{eq:wave_reduced_timoshenko_LRA} must satisfy,
\begin{equation}\label{eq:bloch_theorem_original}
A_{i,n}(x) = A_{i,n-1}(x-L)\, e^{jLq_i}
\, ,
\end{equation}
where $q_i$ are the wavenumbers associated to the acoustic metamaterial as a whole. They should not be mistaken with $k_i$ which are defined solely in the beam. In order to avoid any ambiguity, $q_i$ are often referred to as Bloch wavenumbers. By injecting equation~\eqref{eq:modal_timoshenko} into~\eqref{eq:bloch_theorem_original}, a matrix form can be derived as, 
\begin{equation}\label{eq:bloch_theorem_derived}
\mathbf{C_n} = \mathbf{Q}\, \mathbf{C_{n-1}}
\, ,
\end{equation}
\begin{equation}\label{eq:matrix_H}
\mathbf{Q} = 
\begin{bmatrix} 
e^{jLq_1} & 0 & 0 & 0  \\
0 & e^{jLq_2} & 0 & 0  \\
0 & 0 & e^{jLq_3} & 0  \\
0 & 0 & 0 & e^{jLq_4}  \\
\end{bmatrix}
\, .
\end{equation}
\\
By identification between equations~\eqref{eq:matrix_boundary} and~\eqref{eq:bloch_theorem_derived}, a classical eigenvalue formulation is obtained where $\lambda$ are eigenvalues and $\mathbf{I}$ the identity matrix, such as,
\begin{equation}\label{eq:floquet_bloch}
\lvert \mathbf{X^{-1}} \mathbf{Y} - \lambda \mathbf{I} \rvert = 0
\, .
\end{equation}

\subsection{Acoustic band structure}

This was solved numerically for frequencies, $\omega$, relevant to touch. The propagating modes were obtained by taking the real part of reduced, i.e. dimensionless, Bloch wavenumbers, $q_i L$, defined by,
\begin{equation}\label{eq:real_Qi}
\operatorname{Re}\,(q_iL) = \arctan{ \bigg[ \frac{\operatorname{Im}(\lambda_i)}{\operatorname{Re}(\lambda_i)}\bigg] }
\, .
\end{equation}
It gives the theoretical dispersion diagrams in Fig.~2.B, which delineate a complete band gap. Conversely, the evanescent field characterized by an exponentially decreasing amplitude yields branches within the band gap, obtained by taking the imaginary part of reduced Bloch wavenumbers, $q_i L$, defined by,
\begin{equation}\label{eq:imag_Qi}
\operatorname{Im}\,(q_iL) = -\sqrt{\operatorname{Im}^2(\lambda_i) + \operatorname{Re}^2(\lambda_i)}
\, .
\end{equation}
Numerical evaluations were conducted with dimensions $b=\SI{10}{\milli\meter}$, $L=\SI{10}{\milli\meter}$, assuming linear elasticity of the FR-4 substrate with material properties $\rho=\SI[inter-unit-product = \!\cdot\!]{1.9}{\gram\per\centi\meter\cubed}$\!, $E=\SI{22}{\giga\pascal}$, and $\nu=0.15$. Its thickness, $h$, was optimized within a range commonly recommended by PCB manufacturers, from \SI{0.8}{\milli\meter} to \SI{2.0}{\milli\meter}. Its effect on the band gap width, $\eta$, is illustrated in Supplementary Fig.~\ref{fig:supplementary_theory}.B.

\subsection{Numerical study of Lamb waves}

To quantify the subwavelength capabilities of metamaterials, they are usually compared to reference samples. Here, the reference is a FR-4 beam devoid of local resonators. The wavelength of Lamb waves in this reference was found by solving an eigenvalue problem using finite element analysis in COMSOL Multiphysics 5.5. The FR-4 beam, with dimensions $b=\SI{10}{\milli\meter}$ and $h=\SI{1.6}{\milli\meter}$, was lengthened to $L=\SI{1}{\meter}$ to ensure sufficient resolution for wavelength determination. It was fixed at both ends and subjected to a \SI{205}{\hertz} out-of-plane sinusoidal excitation on its largest face. To accurately model bending, the beam was meshed with $1.2 \times 10^5$ quadratic elements. The material properties were consistent with those used in previous numerical evaluations. The results showed a wavelength $\lambda=\SI{220}{\milli\meter}$ for the asymmetric A0 mode.

\subsection{Experimental band gap evaluation}

Left-hand side transmission coefficients, for $n\in[1,5]$, are presented in Fig.~2.D. These are extended to the right-hand side,  for $n\in[7,11]$, in Supplementary Fig.~\ref{fig:supplementary_band_gap}.A. A band gap ranging from \SI{163}{\hertz} to \SI{290}{\hertz} was observed, which closely matches the left-hand side results, with only a $+2.1\%$ difference in the upper bound. The effect of boundary conditions is illustrated in Fig.~2.G for $n=2$ and extended to the remaining unit cells in Supplementary Fig.~\ref{fig:supplementary_band_gap}.B. Similar conclusions apply as the band gap remains the only region unaffected by changes in boundary impedance. Outside the band gap, the variations of transmission coefficients due to changes in boundary impedance were more pronounced for cells near the boundaries. This underscores the critical role of metamaterials in controlling wave propagation near physical boundaries.

\begin{figure}[!h]
\centering
\includegraphics[width=180mm]{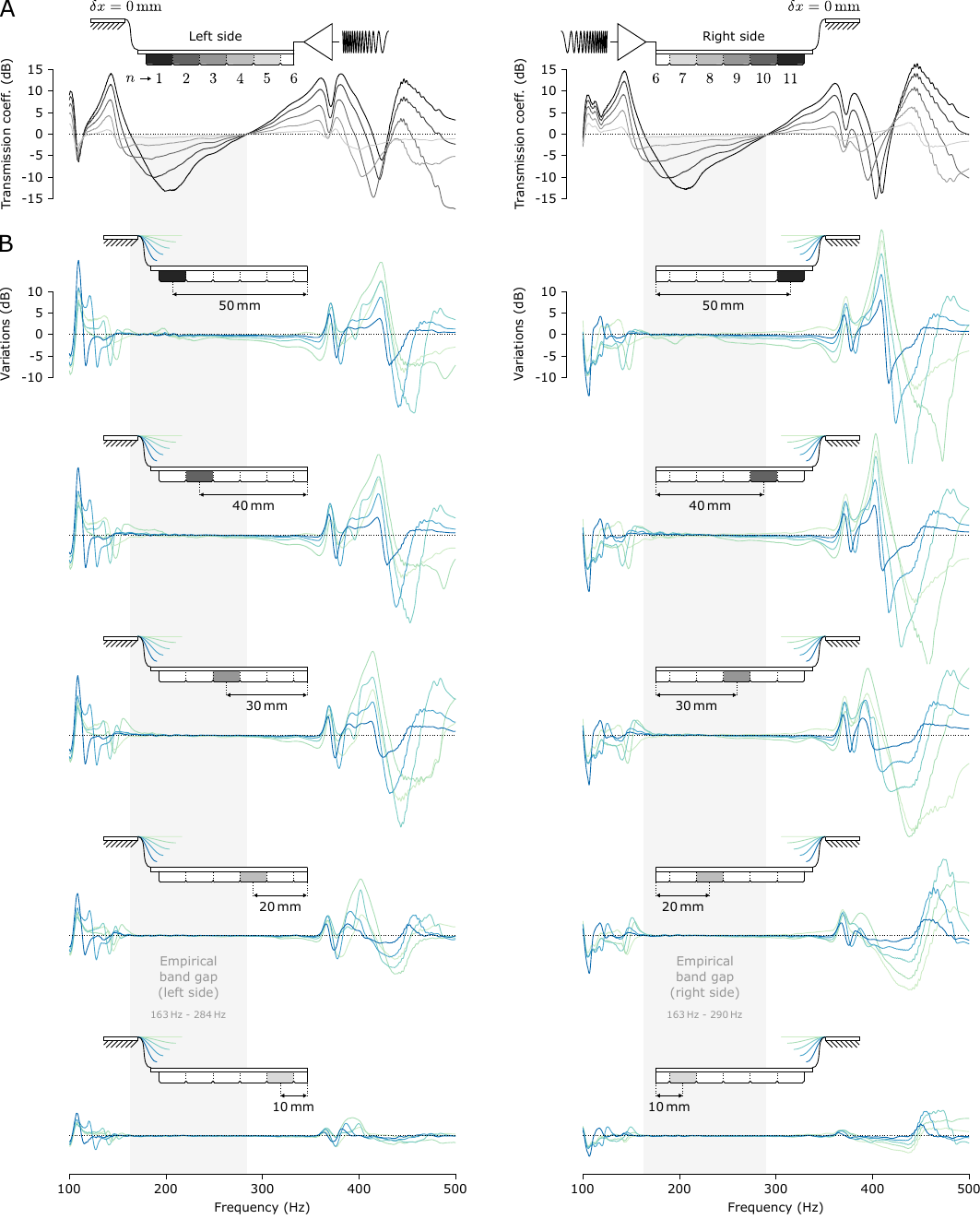}
\captionsetup{width=180mm}
\caption{\scriptsize{\textbf{Detailed band gap analysis.} \textbf{A.} Transmission coefficients on both the left-hand and right-hand sides relative to the central unit cell, $n=6$. \textbf{B.} Corresponding variations in transmission coefficients for increasing levels of flexure tension relative to the reference $\delta x=\SI{0}{\milli\meter}$.}}
\label{fig:supplementary_band_gap}
\end{figure}

\newpage
\section{Prototype}

\subsection{Flexure-based tensioning system}

Both extremities of the metamaterial were suspended by thin flexures made from a \SI{100}{\micro\meter}-thick polyimide film (Kapton, DuPont). These flexures had minimal stiffness and could be approximated as membrane elements. As shown in Supplementary Fig.~\ref{fig:supplementary_flexure}.A and Fig.~\ref{fig:supplementary_electronics}.B, a pop-up tensioning mechanism was designed with the rigid-flex PCB to adjust the tension via a linear motion, $\delta x$. With no tension, $\delta x=0$, the system approximated near free-free boundary conditions. The stiffness of these boundaries increased monotonically with tension, reaching its maximum at $\delta x=\SI{10}{\milli\meter}$.

\begin{figure}[!h]
\centering
\includegraphics[width=88mm]{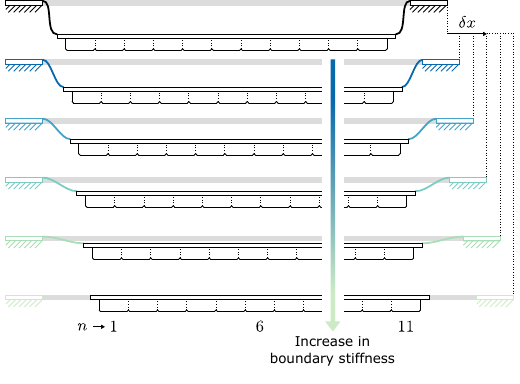}
\captionsetup{width=88mm}
\caption{\scriptsize{\textbf{Flexure tensioning mechanism.} Side view of the metamaterial suspended by thin flexures at its extremities. Their tension is adjusted by the linear setting $\delta x$. The deformation of the flexures is shown at six different tension levels, ranging from \SI{0}{\milli\meter} to \SI{10}{\milli\meter} in steps of \SI{2}{\milli\meter}.}}
\label{fig:supplementary_flexure}
\end{figure}

\subsection{Vibrotactile carrier windowing}

To maximise energy within the band gap and prevent spectral leakage during metamaterial reconfiguration, the sinusoidal carrier was windowed by a Slepian sequence~\cite{Slepian-78}, also known as a discrete prolate spheroidal sequence (DPSS). This sequence is defined in the time domain by a series of functions $w_k(t)$ and in the frequency domain by the corresponding eigenfunctions $W_k(\omega)$, which satisfy the following integral equation,
\begin{equation}\label{eq:dpss_window}
\int_{-\omega_c}^{\,\omega_c} \frac{\sin[\pi N (\omega-\nu)]}{\pi(\omega-\nu)} W_{\!k}(\nu) \, d\nu = \lambda_k W_{\!k}(\omega)
\, ,
\end{equation}
where $\omega_c \in \mathbb{R}$ is the half-bandwidth, $N \in \mathbb{N}$ is the window length, and $\lambda_k$ are the eigenvalues with $k\in[0,N-1]$ being the sequence order. Only the zero-order sequence, corresponding to the largest eigenvalue $\lambda_0$, was used. A narrow half-bandwidth, $f_c = \omega_c/2\pi = \SI{2.5}{\hertz}$, was chosen to effectively concentrate energy within the main spectral lobe, thereby reducing leakage. A comparison with conventional Gaussian windows is given in Supplementary Fig.~\ref{fig:supplementary_windows}.A. For an equivalent main lobe width, a DPSS window exhibits spectral side lobes approximately \SI{20}{\deci\bel} lower than a Gaussian window with a standard deviation $\sigma=0.20N$, as evident from the Fourier transform in Supplementary Fig.~\ref{fig:supplementary_windows}.B. Additionally, a DPSS window with comparable side lobes possesses a narrower main lobe than a Gaussian window with a standard deviation $\sigma=0.15N$, as shown in Supplementary Fig.~\ref{fig:supplementary_windows}.C. These two scenarios clearly demonstrate the superior performance of DPSS windows compared to Gaussian windows.

\begin{figure}[!h]
\centering
\includegraphics[width=180mm]{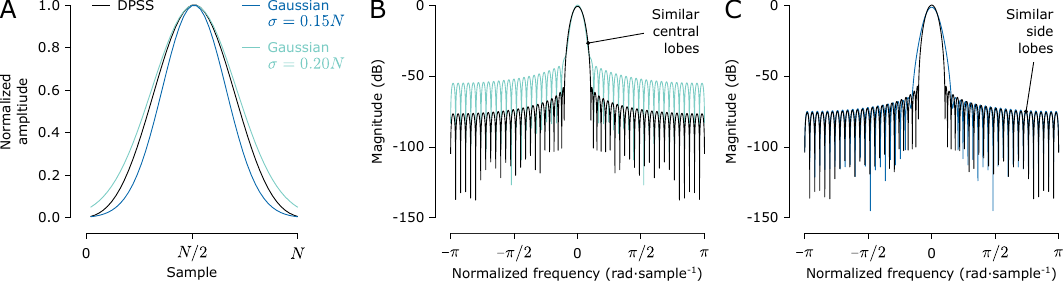}
\caption{\scriptsize{\textbf{Window functions.} \textbf{A.} Normalized amplitude of windows of length $N$, including a DPSS window and two Gaussian windows with standard deviations $\sigma=0.15N$ and $\sigma=0.20N$. \textbf{B.} Fast Fourier transform (FFT) comparison of a DPSS window and a Gaussian window with $\sigma=0.20N$, showing similar central lobes. \textbf{C.} Fast Fourier transform (FFT) comparison of a DPSS widow and a Gaussian window with $\sigma=0.15N$, showing similar side lobe attenuation.}}
\label{fig:supplementary_windows}
\end{figure}

\newpage
\subsection{Embedded electronics}

The embedded electronics in the prototype primarily consist of in a 32-bit microcontroller board (Teensy 4.0, PRJC), H-bridge MOSFET drivers (DRV8837, Texas Instruments), analog accelerometers (ADXL335, Analog Devices), and a buck converter (Okami OKL, Murata), as shown in the schematics in Supplementary Fig.~\ref{fig:supplementary_electronics}.A. 
These components were compactly laid out on a two-layer rigid-flex PCB, as illustrated in Supplementary Fig.~\ref{fig:supplementary_electronics}.B to Fig.~\ref{fig:supplementary_electronics}.D.

\begin{figure}[!h]
\centering
\includegraphics[width=180mm]{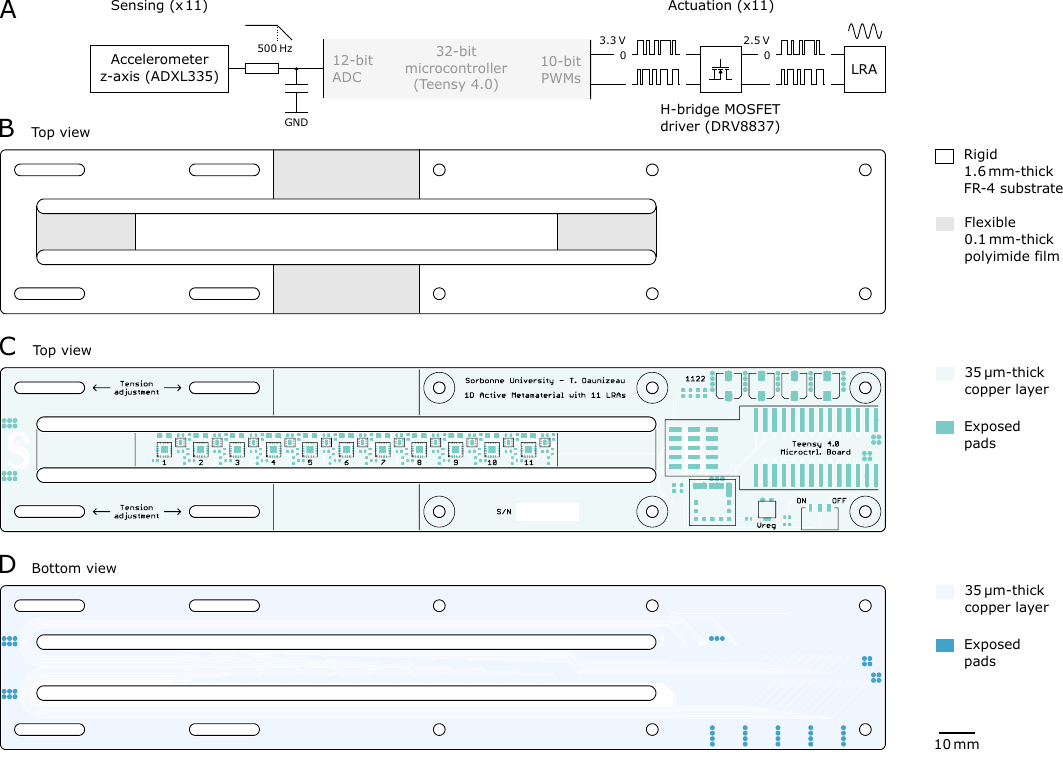}
\captionsetup{width=180mm}
\caption{\scriptsize{\textbf{Embedded electronics.} \textbf{A.} Simplified schematic of the sensing and driving elements in a single cell. \textbf{B.} Top view of the rigid-flex PCB, showing both rigid and flexible parts. \textbf{C.} Layout of the top face. \textbf{D.} Layout of the bottom face.}}
\label{fig:supplementary_electronics}
\end{figure}

\newpage
\section{Supplementary movies}

\subsection{Movie M1}

This movie demonstrates the operating principles of the pop-up structure designed to adjust the tension of the flexures using a rigid-flex PCB. It also illustrates how the parameter $\delta x$ can be manually set.

\subsection{Movie M2}

This movie shows real-time recordings of the RMS acceleration field as 3-bit counting from zero to seven is performed on the metamaterial. The counting is shown at a progressively slower rate, with the time intervals between each number decreasing from \SI{50}{\milli\second} to \SI{5}{\milli\second} in \SI{5}{\milli\second} increments, as illustrated in Fig.~3.I.

\subsection{Movie M3}

This movie shows real-time recordings of the RMS acceleration field as a vibration spot is guided at a constant velocity along the metamaterial. Waveguiding is demonstrated at progressively increasing velocities, $\dot{x}$, ranging from $\SI[inter-unit-product=\!\cdot\!]{100}{\milli\meter\per\second}$\! to $\SI[inter-unit-product=\!\cdot\!]{300}{\milli\meter\per\second}$\! in $\SI[inter-unit-product=\!\cdot\!]{50}{\milli\meter\per\second}$\! increments, as shown in Fig.~5.C.

\subsection{Movie M4}

This movie shows real-time recordings of the RMS acceleration field as a vibration spot follows a sinusoidal reference path, sweeping lengthwise across the metamaterial at \SI{3}{\hertz}, as shown in Fig.~5.E.

\newpage
\bibliography{2024_TD_SH_DG_VH_supplementary}
